\documentclass{amsart}

\usepackage{zi4} 
\usepackage{hyperref}
\usepackage{cite}
\usepackage{booktabs}
\usepackage{amsmath}

\newcommand{\Lisp}{\texttt{Lisp}}
\newcommand{\Clojure}{\texttt{Clojure}}
\newcommand{\Java}{\texttt{Java}}

\newcommand{\code}[1]{\texttt{#1}} 

\usepackage{fancyvrb,newverbs,xcolor}
\definecolor{cverbbg}{gray}{0.95}
\newenvironment{CODE}
 {\SaveVerbatim{cverb}}
 {\endSaveVerbatim
  \flushleft\fboxrule=0pt\fboxsep=.5em
  \colorbox{cverbbg}{%
    \makebox[\dimexpr\linewidth-2\fboxsep][l]{\BUseVerbatim{cverb}}%
  }
  \endflushleft
}

\title{Declarativeness: the work done by something else}

\author{Attila Egri-Nagy}

\address{
Mathematics and Natural Sciences\\Akita International University, Japan\\
\url{egri-nagy@aiu.ac.jp}\\ \url{www.egri-nagy.hu}}

\begin{document}
\maketitle

\begin{abstract}
  Being declarative means that we do computer programming on higher
  levels of abstraction.
  This vague definition identifies declarativeness with the act of
  ignoring details, but it is  a special case of abstraction.
  The unspecified part is some
  computational work.

Automating computations and offloading mental processing are essentially the same concept, which is fundamental for both computational and mathematical thinking.
  This shows that declarativeness is not just a particular style, but it is the  core idea of programming.

  Here we demonstrate this argument and examine its consequences for
  teaching by a systematic study of coding examples from an
  introductory programming course. The chosen language is \Clojure,
  as it is proven to be accessible for novices.
\end{abstract}

\section{Introduction}
`Declarative' is a word we use often and it makes a good sense in
everyday conversation in software development.
For instance, ``\ldots we like to write our programs in such a way that
the code looks like a description of the solution;'' \cite{fogus2014joy}.
Declarative programming is often associated with functional and logic
programming.
Adding database query and configuration management languages
to the list of declarative programming languages would sufficiently cover the meaning.
However, if we try to define it precisely, the concept appears to be elusive.

The legendary SICP \cite{SICP96} mentions ``\ldots the distinction between declarative knowledge and imperative knowledge. In mathematics we are usually concerned with declarative (what is) descriptions,
whereas in computer science we are usually concerned with imperative (how to) descriptions.''
This way `declarativeness' becomes the key difference between mathematics and programming. How is it possible then  talk about programming becoming increasingly more declarative? Is programming becoming more mathematical?

Logic programming refines the what and how distinction. By Kowalski's
equation \emph{algorithm=logic+control}, declarative programming
should deal with the logic part only and leave the control flow part
out. More precisely, a program is a theory, and computation is a
deduction from that theory \cite{Lloyd94}.

Other definitive sources on programming
(e.g.~\cite{knuth2011art,harper2016practical}) simply do not mention the world
declarative, suggesting that it is not an essential concept, or it is
merely a synonym for another important notion.

The vague definition of working on a higher level of abstraction hints that declarative is a relative notion, not
an absolute one. Adding two numbers can be thought as a typical
example of imperative style, but from the perspective of machine code
it is very declarative since we do not specify how to add those
numbers together.
Thus, the level of declarativeness indicates where we are in the hierarchy of abstractions, and the history of programming can be viewed as moving upwards on this abstraction ladder.

It seems that language development has recently reached a sweet spot on the hierarchy of abstractions. A beginner programmer does not have to understand the inner workings of computers, and also does not need to master advanced abstractions. There are of course excellent books for these topics, including computer architecture \cite{petzold2000code}, lambda calculus \cite{LambdaCalculus}, category theory \cite{lawvere2009conceptual, cheng2015cakes, spivak2014category}. However, these require a full course on their own.
Luckily, nowadays a beginner can start programming on a natural level.
What is natural for beginners? It is functional programming, since it is on the familiar level of mathematics that is taught in compulsory math classes.
This is at odds with a clear separation of mathematics and computing.
Indeed, here we argue that unifying computational and mathematical
thinking is possible.
Moreover, it may be the crucial next step in education.

Why is it important to discuss philosophical concepts for computer programming? There is a growing sense of the hidden philosophical assumptions that block or enable the learning of programming, or successfully adapt new technologies. Therefore, investigations in the history and philosophy of computing may be useful \cite{tedre2014science}.
A particularly painful example is  the continued usage of shared mutable sate in parallel concurrent programming.

\subsection{Structure of the paper}
In this paper we define declarativeness by making its implicit
assumption explicit: there is always some automation involved.
Section \ref{Section:declarativeness} defines this manual/automated separation
pattern, which will be used throughout the paper.
Section \ref{Section:Clojure} briefly introduces the \Clojure~language \cite{hickey2008,clojure,halloway2012programming, emerick2012clojure,fogus2014joy}. The core language is small enough that readers with no prior knowledge of \Clojure~can follow the content easily.
Section \ref{Section:Introductory} will analyze examples of
declarative programming accessible for beginners.
Section \ref{Section:Advanced} will go through more advanced examples.

\section{Declarative = the work is done by something else}
\label{Section:declarativeness}

What is the essential idea of declarative programming? Ideas are often
understood in terms of cognitive metaphors, even in software
engineering and computer science \cite{videla2017}.
The leading metaphor here will be automation.
Declarativeness can be captured by separating the work that needs to
be done and the work which is done by a general mechanism, the
automated part.
We will describe examples of declarative programming (and thinking
in general) with tables separating the manual and the automated part
in tasks.
\begin{center}
  \begin{tabular}{lc}
   \toprule
    \multicolumn{2}{c}{\textbf{Cognitive task}}\\
  \midrule
  \textbf{manual} & work we have to do \\
  \textbf{automated} &  work done by a general mechanism \\
  \bottomrule
\end{tabular}
\end{center}

\subsection{Core idea of computation}

If declarativeness is the position on a scale of abstractions, then it
is the amount of detail we need to deal with.
In the context of computation, these are subtasks, i.e.~computational work.
Therefore, in declarative programming the work is done by something else.
Putting it this way, a lot becomes declarative. The whole idea of
using computers can be understood as such.
What we now call programming was described as \emph{automatic programming} in the early days of software engineering \cite{Daylight2012}.

People often say that a declarative piece of code looks like `magic'.
The work done at a lower hierarchical level of computing stack makes it so.
Repeating this relation several times is exactly the magic of digital computation.
As we go down deeper in the computing stack, the work is always done on the level below, bottoming out in simple physical mechanisms that are only capable of processing bit sequences.
\begin{center}
  \begin{tabular}{lc}
   \toprule
    \multicolumn{2}{c}{\textbf{Digital Computation}}\\
  \midrule
  \textbf{manual} & write programs \\
  \textbf{automated} &  a runtime executing our programs \\
  \bottomrule
\end{tabular}
\end{center}

\subsection{Mathematics as declarative knowledge}

One thing is to know what $\sqrt{2}$ is. It is the number that gives $2$ when multiplied by itself.
It is a different thing to know how to obtain its actual value. Here's a method: start with a guess, let's say $1$, divide $2$ by this number $2/1=2$, and take their average $\frac{2+1}{2}=1.5$. By repeating this process we can calculate arbitrary many digits of $\sqrt{2}$.
The very act of denoting a number by the symbol $\sqrt{2}$ is
declarative, as it assumes that the number is available for us (which
would of course require infinite amount of calculation).
It is an infinitely efficient work saving device.
The history is fitting since Turing's original paper was concerned with computing real numbers \cite{Turing1936, AnnotatedTuring2008}.
\begin{center}
  \begin{tabular}{lc}
   \toprule
    \multicolumn{2}{c}{\textbf{Working with real numbers}}\\
  \midrule
  \textbf{manual} & symbolic notation \\
  \textbf{automated} &  algorithms for finding/approximating numerical
    values\\
  \bottomrule
\end{tabular}
\end{center}

Also, an equation can be (and should be when learning math properly)
conceptualized as description of its solution set. This becomes
obvious in analytic geometry, where we the solution sets are
visualized in space. We can talk about the equation of a circle, or
a line as its declarative description.

In differential calculus, the derivation rules compress the work of calculating limits of functions into symbol manipulation formulas. Explicitly pointing this out to students helps them to appreciate more the mathematical results. The derivation rules are work-saving devices, rather than just a set of rules to be memorized.
\begin{center}
  \begin{tabular}{lc}
   \toprule
    \multicolumn{2}{c}{\textbf{Derivation of real functions}}\\
  \midrule
  \textbf{manual} & symbolic manipulation of algebraic formulas using derivation rules \\
  \textbf{automated} &  limit calculation of real functions
    values\\
  \bottomrule
\end{tabular}
\end{center}
\subsection{Unification of mathematical and computational thinking}
In programming it is the reliable code, the well-tested library where we can delegate the work. In mathematics the previously proved theorems play the same role. 
From this general perspective computer programming and mathematical
thinking are very similar. Their common core idea can be summarized easily.
\begin{quote}
  \emph{We use a formal system to lighten the cognitive load or completely offload our thinking processes whenever possible.}
  \end{quote}

  Trying to save work is a natural engineering idea, which applies to the physical world as well. Here we talk about special cases where this is done by and/or through a formal system, a system of symbols and rules for manipulating them.
In the context of formal systems, saving work can be advantageous in
several ways.

\begin{enumerate}
  \item efficiency, to save cognitive effort
  \item promote understanding, increase readability
  \item minimize moving parts to prevent errors
\end{enumerate}

Efficiency is not just about execution speed. It can be decisive between
possible and impossible. Without special training we cannot multiply
large numbers in our head, but using an external representation
(e.g.~pen and paper) it is a routine exercise.

Our understanding is limited by our cognitive constraints. For
instance the number of items we can think of at the same time. The
less pieces we have, the easier to deal with.

Minimizing the number of parts also has an effect on the reliability
of the software product.
The less code we write the less chances we have to make a mistake.

\section{The \Clojure~language}
\label{Section:Clojure}
\Clojure~is a dynamic, general-purpose functional programming language with a special focus on immutable data structures \cite{hickey2008,clojure,halloway2012programming, emerick2012clojure,fogus2014joy}.
Its design makes a strong statement about concurrency and parallelism.
The language is a suggested solution for the difficult problems in concurrency in \Java \cite{JavaConcurrency}, where the relatively low level features of the language make concurrency a formidable problem.
By using persistent and immutable data structures shared memory concurrent processes are easier to handle.

\Clojure~belongs to the \Lisp~family of languages \cite{mccarthy1960}. As such, the core syntax is very simple. Function calls are written as lists. For instance, the list  \code{(f x y)} is evaluated by treating the first element of the list as a function and the remaining items as arguments of the function, corresponding to the mathematical notation $f(x,y)$.
Functions can have different arities. Arithmetic operators are also in the form of a function call.
\begin{CODE}
(+ 1 2 3)
6
(+ 1 7)
8
(+ 2)
2
(+)
0
\end{CODE}
The last constant function's value comes from abstract algebra: 0 is the additive identity, the neutral element).

Some \emph{special forms} (e.g.~conditionals, symbol bindings) have different evaluation patterns, but they all have the same list form.
Function definitions are also special forms.
\begin{CODE}
(fn [x] (* x x))
\end{CODE}
This is a function literal for squaring. It is an anonymous (lambda) function. We can bind it to the name \code{square} at the same time as defining the function.
\begin{CODE}
(defn square [x] (* x x))
\end{CODE}

As a \Lisp~language, the code is represented as lists. Thus the source code coincides with  the abstract syntax tree, a property known as \emph{homoiconicity}.
\Clojure~style is closer to the Programs = Data Structures model, than to the imperative style of Algorithm + Data structures = Programs.
For the same reasons, the metaprogramming facilities are powerful in \Clojure, making it possible to use programming techniques from several different paradigms.

A distinctive feature of \Clojure~is the addition of extra fundamental data structures. Beyond the lists, it has \emph{vectors} (indexable sequential collection, e.g.~\code{[1 2 3 4 5]}, also used for defining function arguments); \emph{hash-maps} (anything to anything hashtables, taking over the roles of objects, e.g.~\code{\{:name Arthur :age 42\}}); and \emph{hash-sets} (hash-maps mapping each key to itself, e.g.~\code{\#\{1 2\}}); all immutable with performance guarantees (persistence).
These all behave like mappings, so indeed, they are also functions.

The simplicity of the core language makes it quite readable.
Therefore, it is possible to illustrate the techniques with relevant source code examples.
This eliminates the need for pseudo-code level descriptions.

\section{Declarative style in introductory programming}
\label{Section:Introductory}

If being relatively more declarative is about going up and down the
abstraction scale, then what is the most natural level of declarativeness?
Of course, the amount of abstraction one can comfortably deal with depends on education and personal development.
But a baseline is easy to establish.
The compulsory part of our math education is
centered around algebraic calculations and real valued functions.
While many feel that there is no  immediate applicability of this knowledge, it is a great entry point for programming.
Functional programming is on the human scale, so it is the easiest to start with for beginner programmers.

In this section we have examples taken from a course designed Liberal Arts students, assuming no background knowledge in programming. More details about the curriculum can be found on the course's website \url{https://egri-nagy.github.io/popbook/}. 
\subsection{Functional collection processing}

The core of an introductory functional programming course can be
defined around the higher order functions \code{map}, \code{filter} and \code{reduce}.
They automate collection processing: \code{map} applies the same function to all elements of a collection, \code{filter} selects elements from a collection satisfying a predicate, and \code{reduce} produces a single result (which might be another collection).
\begin{center}
  \begin{tabular}{lc}
   \toprule
    \multicolumn{2}{c}{\textbf{Functional collection processing
    (\code{map}, \code{filter}, \code{reduce})}}\\
  \midrule
  \textbf{manual} & code for dealing with an element of a collection \\
  \textbf{automated} &  the mechanism of processing a collection \\
  \bottomrule
\end{tabular}
\end{center}

The Collatz conjecture is interesting open problem in number theory, which is easy to state but seems to be an immensely difficult statement to prove \cite{lagarias2010ultimate}.
The conjecture states that the following simple function when iterated starting from a positive integer always end up in a cycle containing 1.
$$\text{collatz}(x)=\begin{cases}
3x+1 & \text{if $x$ is odd } \\
\dfrac{x}{2} & \text{if $x$ is even}
\end{cases} $$
\noindent Due to the lack of regularity in the behaviour of the iterated function, it is a nice problem to explore with computational means.

\emph{What number between 1 and  1000 produces the longest sequence?}
The numerical function is just a translation of the mathematical definition.
\begin{CODE}
(defn collatz [n]
  (if (even? n)
      (/ n 2)
      (inc (* n 3))))
\end{CODE}
Finding the length of the Collatz iteration from a particular number $n$, i.e.~the number of iterations needed to reach 1. With lazy evaluation and higher order functions this is simple task.
\begin{CODE}
(defn c-length [n]
  (count (take-while (fn [x] (not= 1 x))
                     (iterate collatz n))))
\end{CODE}
We count the number of elements in the list built from the consecutive values of the iterated Collatz function.
Once we can calculate Collatz length of a number, we can map this on the sequence of numbers, and find the maximum value.
\begin{CODE}
(apply max
       (map c-length (range 1 1001)))
178
\end{CODE}
The \code{apply} higher order function call is needed since \code{max} is a multi-arity function that is called with a single collection argument. \code{(apply f coll)} is essentially the function \code{f} called with arguments coming from the sequential collection \code{coll}.

Knowing the maximum length, we can find the number that produces that.
\begin{CODE}
(filter (fn [x] (= 178 (c-length x)))
        (range 1 1001))
(871)
\end{CODE}
There happens to be only one such value.
The solution is somewhat unsatisfactory as we need to go through the numbers twice.
With \code{reduce} we can rectify this.
\begin{CODE}
(reduce
  (fn [v n] (let [l (c-length n)]
              (if (> l (first v)) [l n] v)))
  [0 0]
  (range 1 1001))
[178 871]
\end{CODE}
This code does more, but it is also a longer one.
We can transform this solution  by introducing \code{max-key}, which
automates the above steps: finds a number that produces the highest
value for \code{c-length}.
\begin{CODE}
(apply max-key c-length (range 1 1001))
871
\end{CODE}
This shows that a higher order function can move the code closer to
the declarative ideal.
It is also important for the students that they see the transformation
of the code after a correct solution is obtained. This emphasizes the
gain from automation.

\subsection{Point-free style, tacit programming}
Combinatory logic can be used in place of lambda calculus \cite{LambdaCalculus} (for a popular science description of combinatory logic see \cite{smullyan1985mock}).
Defining a function explicitly by describing how it acts on its
arguments can be replaced by several different techniques.

\begin{itemize}
\item composing functions by \code{comp}
\item preloading arguments by \code{partial}
\item grouping functions to work on same input by \code{juxt}
\item negating logical output value by \code{complement}
\end{itemize}

Here are two different functions for calculating the mean of a
sequential collection of numbers (assumed to be non-empty).

\begin{CODE}
(defn mean
  [nums]
  (let [sum (apply + nums)
        n (count nums)]
    (/ sum n)))
\end{CODE}
This function computes the sum of the numbers, their count, and
divides these two.
\begin{CODE}
(def mean
  (comp (partial apply /)
        (juxt (partial apply +) count)))
\end{CODE}
The point-free version defines the function by composing the partial
application of \code{/} with a function that juxtaposes the sum and
the count of its argument collection.

\begin{center}
  \begin{tabular}{lc}
   \toprule
    \multicolumn{2}{c}{\textbf{Point-free style}}\\
  \midrule
  \textbf{manual} & specifying functions to compose \\
  \textbf{automated} & managing input parameters and return values \\
  \bottomrule
\end{tabular}
\end{center}
It is debatable how much tacitness is good for readability, but there
are examples where it led to great success (e.g.~the UNIX pipelines \cite{UNIX1984}).

\subsection{Destructuring}
\emph{Destructuring} is a convenient way to extract pieces of data from a composite data structure.
For example, we can represent a line defined by two points by a vector of vectors and we would like to compute the slope of the line like \code{(slope [[1 2] [2 4]])}. So we define the function \code{slope}.
\begin{CODE}
(defn slope
  [line]
  (/
    (- (first (first line)) (first (second line)))
    (- (second (first line)) (second (second line)))))
\end{CODE}
Here the coordinate information is extracted on demand, making the actual calculation obscure, littered with the retrieval. We can separate these two.
\begin{CODE}
(defn slope2
  [line]
  (let [p1 (first line)
        p2 (second line)
        x1 (first p1)
        y1 (second p1)
        x2 (first p2)
        y2 (second p2)]
    (/ (- x1 x2) (- y1 y2))))
\end{CODE}
Now the computation is quite clear, it is basically the mathematical formula. However, we have a long list of bindings. Destructuring gets rid of this, by giving the `shape' of the input data in the argument list.
\begin{CODE}
(defn slope3
  [[[x1 y1] [x2 y2]]]
(/ (- x1 x2) (- y1 y2)))
\end{CODE}
This may look like magic first, but it is actually just a simple automation.
It is easy to reveal how it is done.
\begin{CODE}
(destructure '[ [x y] [13 19]])
[vec__1246 [13 19]
 x (clojure.core/nth vec__1246 0 nil)
 y (clojure.core/nth vec__1246 1 nil)]
\end{CODE}
It does exactly the work we did not want to do manually.
\code{destructure} produces a vector of bindings, that is given to \code{let} in a real destructuring situation.
It is also very useful in figuring out what goes wrong in an unsuccessful and complex destructuring attempt.
\begin{center}
  \begin{tabular}{lc}
   \toprule
    \multicolumn{2}{c}{\textbf{Destructuring}}\\
  \midrule
  \textbf{manual} & describing the shape of a data structure \\
  \textbf{automated} & extracting data, local bindings \\
  \bottomrule
\end{tabular}
\end{center}

\section{Declarative style in advanced programming}
\label{Section:Advanced}

Some declarative programming concepts cannot be fit into an
introductory class. This might be due to time constraints, or due to
the fact that the less declarative form is more general. For instance
conditionals can be used for any decision making, but the more elegant
pattern matching may have some limitations and not applicable in all cases.

\subsection{Pattern matching}

Destructuring is just a special case of \emph{pattern matching}.
We can also use it for making decisions based on the structure of the data. 
In \Clojure~there is a core library for pattern matching: \code{core.match} is based
on techniques from \code{OCaml} \cite{2008patternmatching}.
\begin{CODE}
(require '[clojure.core.match :refer [match]])
\end{CODE}
The archetypal example for pattern matching is the Fizzbuzz
game.
\begin{CODE}
(defn fizzbuzz [lim]
 (for [n (range 1 lim)]
  (match [(mod n 3) (mod n 5)]
      [0 0] "FizzBuzz"
      [0 _] "Fizz"
      [_ 0] "Buzz"
      :else n)))
\end{CODE}
Matching replaces conditional statements. When nested, conditionals seem to be difficult to
read and thus error-prone.
\begin{center}
  \begin{tabular}{lc}
   \toprule
    \multicolumn{2}{c}{\textbf{Pattern matching}}\\
  \midrule
  \textbf{manual} & specifying choices based on the structure of data \\
  \textbf{automated} &  control flow by conditional statements \\
  \bottomrule
\end{tabular}
\end{center}

Since the order of the patterns does matter, the programmer still has
to think in terms of the underlying matching process.

\subsection{Logic programming}
Logic programming is considered to be a higher level of declarativeness on two accounts.
It is a generalization of functional programming to relational programming, so in a sense functions can be run backwards.
It is also a more general type of pattern matching. The underlying unification operation can be viewed as two-way pattern matching.

Most programming problems  (without need for user interaction) can be rephrased as search problems.
Moreover, they can be easily described by using first order logic. These two combined lead  to the idea of logic programming.
Its promise is that we only need to specify properties of the required solution.
\begin{center}
  \begin{tabular}{lc}
   \toprule
    \multicolumn{2}{c}{\textbf{Logic programming}}\\
  \midrule
  \textbf{manual} & specifying properties of a solution by clauses\\
  \textbf{automated} &  finding models for the properties by a search algorithm\\
  \bottomrule
\end{tabular}
\end{center}
On the other hand, the most important logic programming language, \code{Prolog}, kept  tools for manipulating the search algorithm. For instance, cutting the search tree or changing the underlying database (see standard references of the language like \cite{clocksin2003prolog,bramer2005logic}).
A more declarative version of logic programming is \emph{relational programming}, which can be viewed as `pure' logic programming free from low-level access, or as a generalization of functional programming where the directionality of computation from function arguments to its result is removed.
This style fits \Lisp-like languages well \cite{friedman2005reasoned,byrd2009relational}, but can be implemented in any language (see \url{minikanren.org} for a list of implementations).

\subsubsection{Relation programming in \Clojure: \code{core.logic}}

The \Lisp~nature (especially its macro system) of \Clojure~
makes it possible to accommodate
different programming paradigms as an additional library.
For instance, the ~\emph{Thinking programs} chapter of \cite{fogus2014joy} describes several solutions for a sudoku solver, starting from a brute force functional solution to one using unification. Here we describe a smaller such transition. The task is a simple combinatorial problem, producing all permutations of (assumed to be unique) elements of a collection.
Here is a classic functional recursive implementation.
\begin{CODE}
(defn permutations
  [coll]
  (if (empty? coll)
    [coll]
    (mapcat (fn [x]
              (map (partial cons x)
                   (permutations (remove (partial = x) coll))))
            coll)))
\end{CODE}
If a collection is empty, then we can immediately return all permutations of its elements, which is the empty collection itself.
If not empty, then for each element we recursively create the collection of all permutations of the remaining elements and include the chosen element in front. These collections have to be flattened (by \code{mapcat}) for each depth on the recursion.
While the recursive style with the higher order functions can be seen as elegant, it is very much a how-to description. When reading the code snippet, one has to imagine the recursive process of building the permutations backwards.

The same problem can be solved with a logic programming approach, offering a more declarative solution. In \Clojure~we can switch paradigm just be including a library.
\begin{CODE}
(require '[clojure.core.logic :as l]
         '[clojure.core.logic.fd :as fd])
\end{CODE}
This piece of code loads the logic engine and a library for dealing with finite domains.
\begin{CODE}
(defn permutations [n]
  (let [p (vec (repeatedly n logic/lvar))
        points  (fd/interval 1 n)]
    (logic/run* [q]
      (logic/== q p)  
      (logic/everyg (fn [x] fd/in x points) p)
      (fd/distinct p))))
  \end{CODE}
  The logic code is wrapped in a normal function definition.
  \code{p} is a vector of $n$ logic variables, and \code{points} is just a finite domain defining their possible values.
  \code{run*} calls the logic engine requesting all solutions in \code{q}, which is the query variable.
  \code{q} is associated with \code{p} by the unification operator \code{==}.
  The last two lines are the description of a solution: \code{everyg} takes a goal (a predicate function) and checks
whether it holds for all elements of a collection, code{distinct} checks for any repeated element.
Thus the code reads as `find all the $n$-tuples with integers from the
interval $[1,n]$ such that the elements of the tuples are all distinct'.

As the above example illustrates, \code{core.logic} is not directly accessible for an absolute beginner. It already requires the understanding of higher order functions and some basic knowledge of \Clojure.

\subsection{SAT-solvers}

Another very interesting case of logic programming is the progress of efficient general purpose solvers for the satisfiability problem \cite{KnuthSAT,biere2009handbook}.
While satisfiability is `the' hard problem, the problem instances coming up in practical applications often have enough structure for finding solutions quickly. 
Thus, if a computational problem can be represented as a conjunctive normal form, then we can rely on the performant search techniques.
\begin{center}
  \begin{tabular}{lc}
   \toprule
    \multicolumn{2}{c}{\textbf{Search problems}}\\
  \midrule
  \textbf{manual} &  describe search problem in CNF\\
  \textbf{automated} & SAT-solver finds solution configurations \\
  \bottomrule
\end{tabular}
\end{center}
Relying on the general solution algorithm (and all of its sealed off optimizations) has another benefit.
For computational projects, where a bespoke search algorithm is needed (e.g.~\cite{T4}), SAT-solvers still can serve as validation methods.
\section{Conclusion}

Here we attempted to clarify the vague notion of declarative programming by interpreting each of its occurrences as \emph{a separation of manual and automated computation}.
We showed that abstraction could almost cover the meaning of
declarative, but it does not necessarily involve the work saving part.
Therefore, the usage of the term declarative programming is justified.
While it cannot readily provide a quantitative measure, it would be beneficial if each use of the term came with a clear indication of what is automated and what needs to be done by the programmer.

This general definition also invites rethinking the similarity and differences of mathematical and computational thinking. For education though, it is decisive to emphasize the similarities. We use a formal systems for offloading our cognitive processes, so both mathematics and programming languages automate our thinking.
\bibliographystyle{plain}
\bibliography{declprog}

\end{document}